# EVALUATION OF MACHINING DISPERSIONS FOR TURNING PROCESS

**Lefebvre Arnaud**
PRISMa, Claude Bernard University – Lyon 1,
17 rue de France, 69627 Villeurbanne, France
Tel : 04 72 65 54 80, arnaud.lefebvre@iutb.univ-lyon1.fr

**Wolff Valéry**
IUT B – GMP, 17 rue de France, 69627 Villeurbanne Cedex,
Tel : 04 72 65 54 80, valery.wolff@iutb.univ-lyon1.fr

## Abstract:

*In order to control the design of product/process, the manufacturing companies are directed more and more towards the installation of tools of simulation and in particular simulation of machining. The simulation of machining aims to simulate the operations of machining by taking into account the dispersions of machining in order to determine the intervals of tolerances and to minimize the manufacturing costs.*

*In this article we propose a model of simulation of dispersions in turning based on the geometrical specifications. Our study is articulated around two trends of development: the first trend relates to the geometrical model. The geometrical model suggested must allow a follow-up of the geometry of the part during the simulation of machining. It is thus a question of carrying out a systematic treatment of the whole dimensioning and tolerancing process while being based on the principles of the ΔL method. One of the limits of the traditional ΔL method resides in the one axis exclusive study of a dimension of a part. For the turned workpieces for example the specification of coaxiality often specified by the tolerancing activity is not taken into account by this method. We also planned to integrate this type of specification in the model of simulation of machining suggested. It is more generally a question of extending the traditional model for better taking into account the multi axis specification of coaxiality and perpendicularity on the turned workpieces.*

*The second trend of our study relates to the widening of the field of application of the model. We propose to extend the field of application of the model by taking into account the modifications of several parameters of the manufacturing process plans, likely to involve variations of dispersions. The use of the design of experiments method makes it possible to quantify the influence of these modifications on dispersions. Integration with the initial model of parameters relating to cutting conditions or the nature of the material – parameters invariants in the initial model - allows on the one hand, to extend the field of application of the model and allows on the other hand to enrich considerably the model by simulation of dispersions of machining. This experimental study based on the model of design of experiments made it possible to confirm the assumptions of invariants retained for the construction of the initial model. Moreover, the influence of several cutting parameters on particular dispersions of machining could be revealed and quantified.*

The stage of simulation of machining allows to simulate the behaviour of the operations of the process plan considering the dispersions to fix the intervals of tolerances and to produce at lower cost.

The first part of this article is devoted to modelling the dispersions in turning. We propose to extend the traditional model of ΔL by taking into account defects of coaxiality and perpendicularity on a turned part.

Experimental models resulting from design of experiments make it possible to quantify the influence of several parameters on dispersions. The detailed analysis of a model of dispersions of machining on Z axis in the second part of this article allows to quantify the variations of dispersions according to the evolution of the cutting speed and the insert type.

## 1 Modelling of dispersions

We call dispersions of machining the geometrical and dimensional variations obtained on a set of real parts for a manufacturing process plan and a given machine-tool. The supposed sources of dispersions have several origins in particular related to controls with the inflexion of the tools, the cutting efforts and the geometrical defects of the machine tool. The rules which characterise each origin of dispersions can be of various nature (normal, bimodal…); nevertheless we make the assumption that the resultant follows a normal distribution.

### 1.1 Extended model suggested

Considering a shouldered part the modelling of the behaviour of a machine-tool [3] is classically approached according to a study of five parameters of Δ*machine* dispersions as shown in the figure 1. These dispersions are classified in two categories: the first category relates to dispersions of setting in position such as ΔO, Δα, $\Delta Z_r$. The second category includes dispersions of machining like $\Delta R_u$ (dispersion of machining according to X axis) and $\Delta Z_u$ (dispersion of machining according to the Z axis)[1].

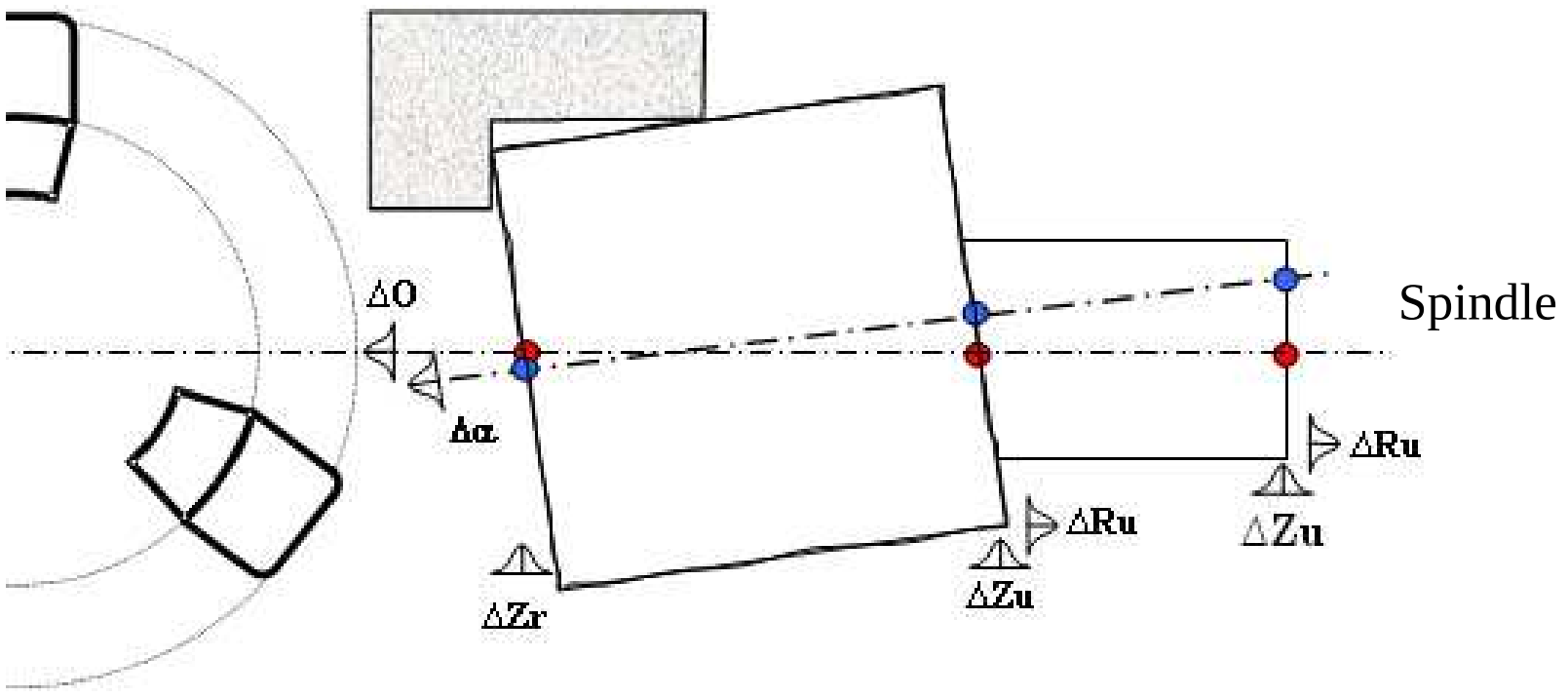

*figure 1 : Modelling of the 5 dispersions in turning*

We use the traditional method of ΔL on X and Z axes to take into account the axial dimensions and the dispersions obtained during machining. The parameters are then indicated by $\Delta Z_u$, $\Delta R_u$, $\Delta Z_r$. We must use new parameters to take into account the geometrical specifications of tolerancing (coaxiality, perpendicularity…): Δα and ΔO [7].

- Δα represents the angular defect of remachining of the part in the jaws in turning,

- ΔO is the defect of concentricity (between the axis of the reference surface and the spindle axis) located at the bottom of the soft jaws,

- $\Delta Z_r$ corresponds to the axial remachining error of the part in the part holder along the Z axis.

The objective of the modelling of this study of machining is to determine relationships characterising the behaviour of the machine-tool in terms of dispersions. These relations are of the type: $\Delta_{machine} = f_i(p_i)$ where $\Delta_{machine}$ is one of the five dispersions and $p_i$ a set of discrete or continuous product/process parameters.

## 1.2 ISO tolerancing

ISO Standards of tolerancing gathered under the term of GPS (Geometrical Products Specifications) provide a complete language to the mechanical engineers [2]. They are adopted today by the manufacturing industry. The geometrical model of dispersions in simulation of machining which we propose [9] takes into account the ISO specifications.

For example in the case of the coaxiality we defined the methods of calculation [7] necessary in order to connect the parameters of our model to ISO specifications.

The coaxiality relates to the relative position of the real axis of the toleranced surface and the reference axis. It never relates to surfaces but always to axes. The definition resulting from the standard and its interpretation is defined figure 2.

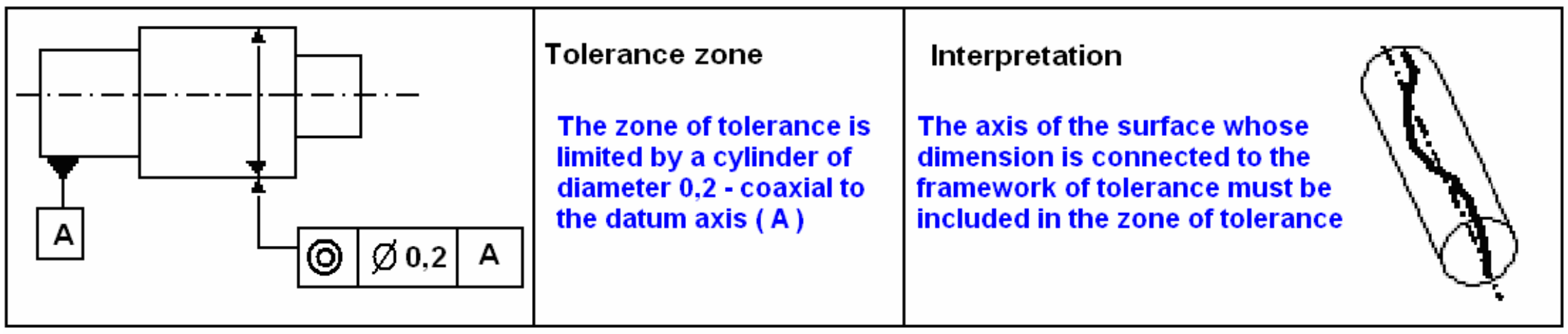

*figure 2 : GPS standards- coaxiality according to ISO 1101*

This definition must be interpreted and calculated most precisely possible to make correspond the measurement obtained on three-dimensional coordinate measuring machine (CMM) with the specifications of the drawing.

## 1.3 Experimental protocol

### 1.3.1 Design of experiments (DOE) parameters

We defined five product/process parameters allowing the evaluation of various dispersions. The table 1 gives the list of the parameters and their associated levels.

| Parameters | | Type | Values | |
|---|---|---|---|---|
| Insert type | N | discrete | P15 (finish) | P35 (rough) |
| Nose radius (mm) | Rε | discrete | 0.40 | 0.80 |
| Cutting speed (m/min) | Vc | continuous | 150 | 280 |
| Material of the machined part | M | discrete | A60 | XC38 |
| Feed rate (mm/turn) | F | continuous | 0.10 | 0.30 |

*table 1: Factors and associated values*

Particular interactions are taken into account to supplement the study (table 2).

| Interactions | |
|---|---|
| Insert type - cutting speed | $N.V_c$ |
| Material – feed rate | M.f |
| Cutting speed - material | $V_c.M$ |
| Insert type – feed rate | N.f |

*table 2: Interactions retained between parameters*

#### 1.3.2 DOE Choice

We use the Taguchi's method [4][5][6] to limit the number of tests to be realised. The table $L_{16}$ ($2^{15}$) table was selected according to the criterion of orthogonality and number of degrees of freedom of the model. For each response of studied dispersion, the order of the experiments as well as the combinations of the parameters are given in table 3.

| N° | Insert type | Nose radius Rε | Cutting speed $V_c$ | Material M | Feed rate f |
|---|---|---|---|---|---|
| 1 | P15 | 0.4 | 150 | A60 | 0.1 |
| 2 | P15 | 0.4 | 150 | XC38 | 0.3 |
| 3 | P15 | 0.4 | 280 | A60 | 0.3 |
| … | … | … | … | … | … |
| 16 | P35 | 0.8 | 280 | XC38 | 0.1 |

*table 3: Table of the experiments (extract)*

## 1.4 Experimental results

#### 1.4.1 Evaluation of the answers

The responses of the experiments are obtained by measuring. The measures allow to calculate the dispersion parameters: ΔO, Δα, $\Delta R_u$, $\Delta Z_r$ and $\Delta Z_u$ (millimetres). A reduced sample of five workpieces is enough. The used Taguchi's method is a standard $L_{16}$ ($2^{15}$) experiments 5 times repeated.

We observe two cases. The response is calculable starting from the standard deviation obtained by the measurements of one or several dimensions:

- Case of the response obtained by the measurement of only one dimension. That relates to dispersions ΔO, Δα, $\Delta R_u$. For example, the relation relating to $\Delta R_u$ is written:

$$\Delta R_u = \left[ \frac{6 \times (\sigma_{diametral})_{sample}}{C_4} \right] / 2 \qquad (1)$$

where $C_4$ is the weighting taken in the statistical table of the reduced samples.

- Case of the response obtained by the measurement of several dimensions $d_i$ (standard deviation noted $\sigma_i$). The variance of required dispersion is related to the sum of the variances of dimensions concerned dimensions. That relates to $\Delta Z_r$ dispersions (or $\Delta Z_u$). The relations used are as follows:

$$\sigma_r = \frac{\sqrt{\sum \pm \sigma_i^{\,2}}}{\sqrt{2}} \text{ and thus } \Delta Z_r = \frac{(\sigma_r)_{sample}}{C_4} \times 6 \qquad (2)$$

The necessary values to calculate the parameters of dispersions (part diameters, points of intersection…) are measured directly on the 80 parts using a coordinate measuring machine. Some of these values are only intermediates parameters.

Each batch of 5 parts allows to calculate the standard deviation of each answer $\Delta Z_r$, $\Delta O$, $\Delta\alpha$, $\Delta R_u$ and $\Delta Z_u$ using CMM measurements. The design of experiments provides in this way 16 values for each studied response.

#### 1.4.2 Analyse and summary of the results

We carried out the analysis of the measurements obtained on the 80 parts of the $L_{16}$ ($2^{15}$) plan definite previously to determine the parameters of manufacture process plan influencing machining dispersions.

The analysis of the variance indicates that a parameter is statistically significant on the response as soon as *p* parameter is higher than 0.05 (Level of confidence higher than 95%).

The R-squared ($R^2$) makes it possible to evaluate the percentage of data explained by the model. The higher the $R^2$ is, the more the model is usable in a predictive mode. A $R^2$ coefficient between ~70 and ~ 90% corresponds to an acceptable model.

## 2 Analysis of dispersions models

The table 4 presents the summary of the results of obtained dispersions. Only are presented in the table the answers for which parameters appear statistically significant. It is thus the case of dispersions $\Delta R_u$, $\Delta Z_{u_dressage}$ and $\Delta Z_{u_en_remontant}$.

Note: One of the characteristics of the model suggested is due to the distinction made on the level of the dispersion in $\Delta Z_{u_dressage}$ and $\Delta Z_{u_en_remontant}$ [7]. Two distinct answers will then be evaluated.

| **Dispersions responses** | **Statistically significant parameters** |
|---|---|
| $\Delta R_u$ | Material – Cutting speed (M Vc)<br>Nose radius (Rε)<br>Material – Feed rate (M f) |
| $\Delta Z_{u_dressage}$ | Cutting speed (Vc), Insert type (N) |
| $\Delta Z_{u_en_remontant}$ | Material – Feed rate (M f) |

*table 4 : Statistically significant parameters*

## 2.1 Responses modelling

For each response of the table 4, the machining dispersion can be represented in the form of a regression function.

The obtained model by linear regression concerning $\Delta R_u$ dispersions is:

$$\begin{aligned} \Delta Ru &= 0.00876824 + 0.0182206 \times N + 0.0318719 \times R\varepsilon + 0.000080 \times Vc + 0.0118769 \times M \\ &+ 0.019411 \times f - 0.0000424 \times N \times Vc \\ &- 0.0287376 \times N \times f \\ &- 0.0001095 \times Vc \times M + 0.0615979 \times M \times f \end{aligned} \tag{3}$$

The obtained model for $\Delta Z_{u_dressage}$ dispersions is:

$$\begin{aligned} \Delta Zu_dressage &= 0.0214865 - 0.00477308 \times N - 0.005125 \times R\varepsilon - 0.00004307 \times Vc \\ &+ 0.0015125 \times M + 0.007 \times f + 0.000013653 \times N \times Vc \end{aligned} \tag{4}$$

And the model obtained for $\Delta Z_{u_en_remontant}$ is:

$$\begin{aligned} \Delta Zu_en_remon\tan t &= -0.00899519 - 0.0151385 \times N + 0.0171875 \times R\varepsilon - 0.0000196 \times Vc \\ &+ 0.0122423 \times M + 0.108 \times f + 0.00004942 \times N \times Vc + 0.069375 \times N \times f + 0.00001538 \times Vc \times M \\ &- 0.128 \times M \times f \end{aligned} \tag{5}$$

These mathematical models have to be used in a predictive mode to evaluate machining dispersions for the various combinations of product/process parameters defined in the table 1 [8].

## 2.2 Representation of $\Delta Z_{u_dressage}$ model

### 2.2.1 Graphical chart

Contrary to the mathematical model resulting from the design of experiments (see §2.1), the representation of the $\Delta Z_{u_dressage}$ dispersions model is limited. Indeed, we propose a representation which deals with statistically significant parameters (see table 4).

In the case of $\Delta Z_{u_dressage}$, the 2 statistically significant parameters are the cutting speed (Vc) and the insert type (N). Then $\Delta Z_{u_dressage} = f($ Vc, N $)$

The represented surface on figure 3 shows the various ranges of variation of dispersions according to the 2 parameters.

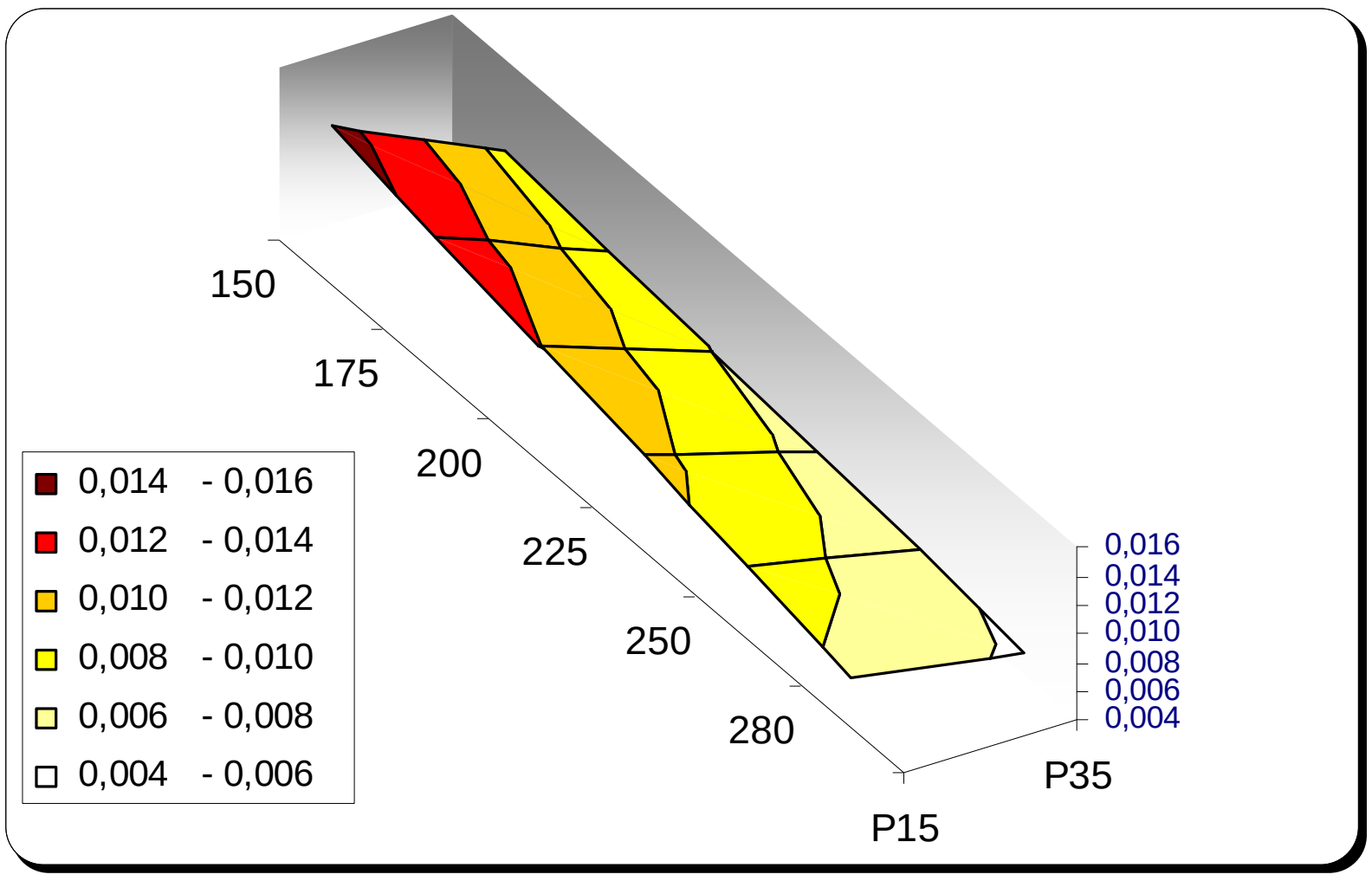

*figure 3 : $\Delta Z_{u_dressage} = f(Vc, N)$*

### 2.2.2 Analyse of dispersions variations

The following graph in figure 4 represents the evolution of $\Delta Z_{u_dressage}$ dispersion according to the cutting speed (Vc) for each of the 2 modal values of the insert type (P15, P35).

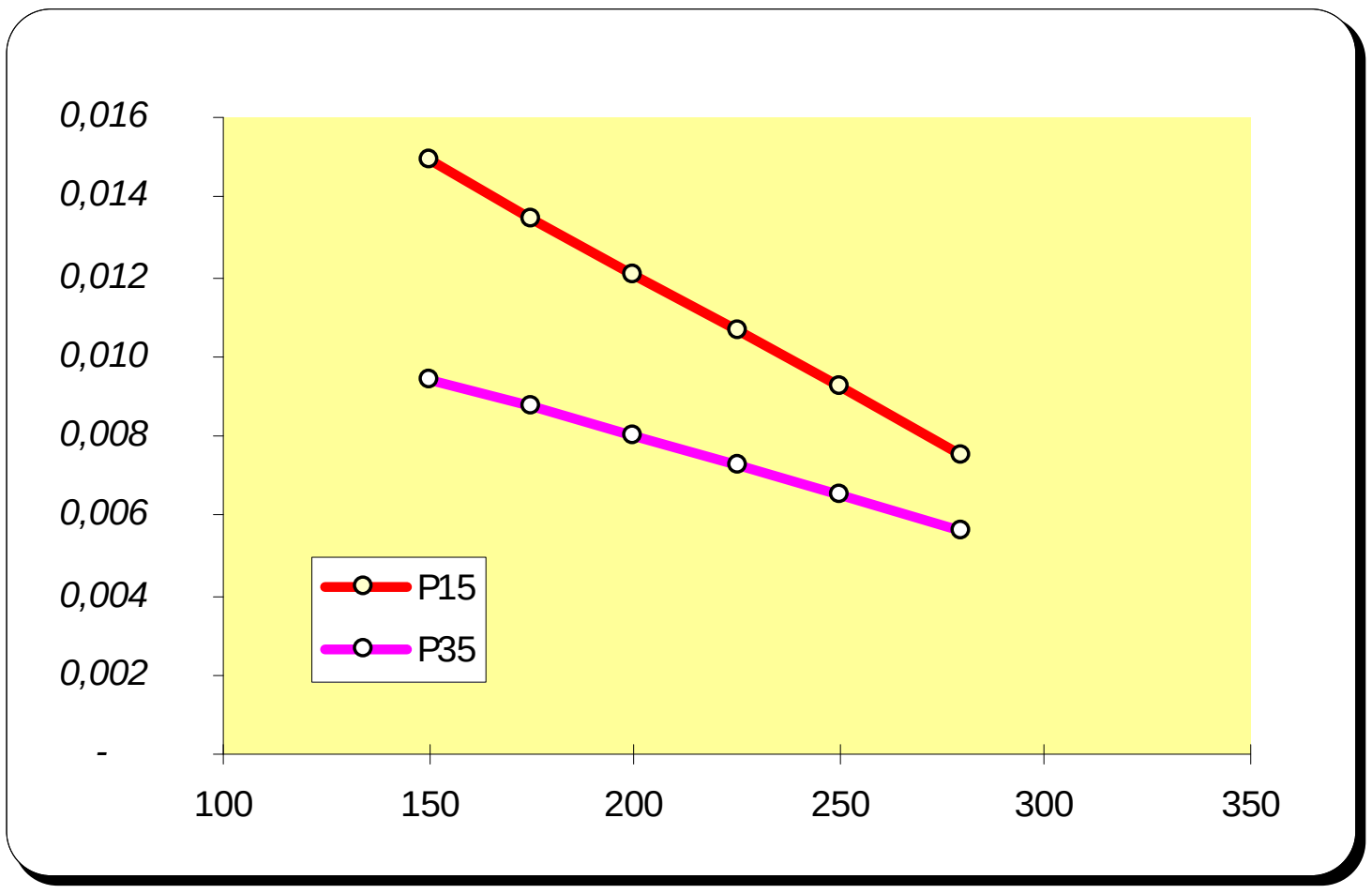

*figure 4 : Cutting speed influence on dispersions*

a) We can see on figure 4 a considerable influence of the cutting speed ($V_c$) on the values of dispersions obtained by the predictive model. Indeed the values obtained for the P15 insert type vary from 16 to 8 µm in the range of variation of the cutting speed. We also observe the same trend for the P35 insert type.

The values of dispersions are lower for the P35 insert type than for the P15 insert type. This is explained by the choice of the ranges of the cutting speed. The chosen cutting speeds for the experimental protocol must follow 3 use rules:

- The cutting speed range (Vc) must be compatible with the manufacturers insert types database.

- The limits of the machine-tool (maximum spindle speed) must be taken into account.

- The construction of a design of experiments requires to be able to use the same ranges of variation speed (Vc) whichever the other parameters values are (material, nose radius, feedrate…).

In the case of the P15 insert type which is a finish insert type the used range is located partly low (minimum speed) of the possible range given by the manufacturer. Preconnized optimal speed is higher than 400 m/min. For the P35 insert type the totality of the range of the preconnized cutting speed was used. The maximum preconnized speed is close to 250 m/min.

b) We observe a convergence of these dispersions for the 2 modal values P15 and P35 when it reaches the maximum values of the chosen range for $V_c$.

For the P15 insert type this is explained by the fact that at the convergence point the used speed (280 m/min) approaches the optimal speed.

In the next future, we will simulate the process planning of a turned part using the predictive models according to the different cutting parameters. We will then validate this predictive models by using a set of machined parts (real parts).

## Conclusion

The simulation of machining and the characterisation of the machine-tools in terms of dispersions of machining are a necessary stage in the control of the costs. In the first part of this article we presented an extended model of dispersions in turning based on multiaxes defects of coaxiality or perpendicularity. It allows thus to take into account a more complete definition of the geometrical model - as well on the dimensional level and on the geometrical level. Moreover, for the five types of selected dispersions, angular dispersions, remachining dispersions, an experimental model was elaborated and gave us the possibility to quantify the influence of several parameters like the type of material or the cutting speed.

In the second part of the article, we clarify an example of dispersions model. We showed in experiments the influence of the cutting speed as well as the insert type. The model which is graphically represented allows to quantify the values of dispersions according to the statistically significant parameters. These results allow on the one hand to enrich considerably the initial model of dispersions in which the process plan parameters were fixed. In addition to the aspect of formalisation of a know-how, they allow to better characterise the machine-tool and thus to improve the quality of the machined parts.

## Références